# Nonlinear optical properties of mono-functional 1,2-dihydro-1,2-methanofullerene[60]-61-carboxylic acid /polymer composites

Hendry Izaac Elim,[a] Jianying Ouyang,[b] Jun He,[a]
Suat Hong Goh,[b] Sing Hai Tang[a] and Wei Ji[a*]

[a] Department of Physics, National University of Singapore,
3 Science Drive 3, Singapore 117543, Singapore
[b] Department of Chemistry, National University of Singapore,
3 Science Drive 3, Singapore 117543, Singapore

**ABSTRACT**

By using nanosecond laser pulses at 532-nm wavelength, we have studied the nonlinear optical properties of composites which consist of mono-functional 1,2-dihydro-1,2-methanofullerene[60]-61-carboxylic acid (FCA) and poly(styrene-co-4-vinylpyridine). The optical limiting performance of FCA itself is poorer than that of its parent $C_{60}$, while FCA incorporated with polystyrene shows better optical limiting responses, with the limiting threshold reduced by about 35%. In addition, the FCA gives slightly stronger photoluminescence emission than that of both $C_{60}$ and FCA/polymer composites. The possible sources for the improvement in the optical limiting are discussed.

---

[*] Corresponding author, email address: phyjiwei@nus.edu.sg

# 1. Introduction

Considerable research efforts have been focused on the applications of [60]fullerene due to their promising use as an excellent optical limiter for eye and sensor protection from high intensity laser pulses [1,2]. Other potentials such as optical switches and a single-$C_{60}$ transistor have been investigated [3,4]. Recently strong magnetic signals have been found in rhombohedral $C_{60}$ polymer (Rh-$C_{60}$), and studies of synthetic route to the $C_{60}H_{30}$ polycyclic aromatic hydrocarbon and its laser-induced conversion into fullerene-$C_{60}$ have also received attention [5,6].

The main challenge that still remains is how to improve the optical limiting and nonlinear optical properties of $C_{60}$. A possibility to tackle this problem is to explore $C_{60}$ derivatives, in which $C_{60}$ is attached by complementary functional groups of some polymer [7-10]. It is expected that the optical properties of $C_{60}$ derivatives change upon functionalization. Sun and his co-workers have investigated a series of methano-$C_{60}$ derivatives and they have found that the mono-functional $C_{60}$ derivatives show similar optical limiting responses to that of parent $C_{60}$ while multi-functional ones give poorer performance [11-13].

In this paper, we present an investigation into the nonlinear optical properties of mono-functional 1,2-dihydro-1,2-methanofullerene[60]-61-carboxylic acid (FCA)/polymers composites. The fluence-dependent transmission measurements conducted with 532-nm, nanosecond laser pulses show that the optical limiting behavior of FCA incorporated with polymer is better than that of $C_{60}$, with the limiting threshold reduced by about 35%. The possible sources for the improvement are analyzed and discussed with Z-scan and photoluminescence (PL) measurements.

**2. Experimental**

Our $C_{60}$ (99.9%) sample was obtained from Beijing University, Beijing, China. The concentration of $C_{60}$ in this experiment was $1.6 \times 10^{-3}$ M. 1,2-dihydro-1,2-methanofullerene [60]-61-carboxylic acid (FCA) was synthesized by the method reported by Isaacs and his co-workers [14,15]. The FCA has been prepared carefully to be [6,6]-closed isomer with 58 $\pi$-electrons. Poly (styrene-co-4-vinylpyridiene) (PSVPy) and polystyrene (PS) were prepared by free-radical copolymerization initiated by an initiator for polymerization (AIBN). The molar percentage of 4-vinylpyridine unit in PSVPy32 was 32% as determined by elemental analysis. $C_{60}$ and FCA were dissolved in toluene and 1,2-dichlorobenzene, respectively, into which appropriate amounts of PSVPy32 and PS were added. Six samples were prepared, as shown in Table 1. Both the structure of FCA and the absorption spectra of the FCA/polymer composites are shown in Fig. 1.

The optical limiting measurements were conducted using linearly polarized nanosecond optical pulses from a Q-switched, frequency-doubled Nd:YAG laser (Spectra Physics DCR3) with pulse duration of 7 ns or an optical parametric oscillator (Spectra Physics MOPO 710) with pulse duration of 5 ns. The spatial distribution of the pulses was nearly Gaussian after passing through a spatial filter. The pulse was divided by a beam splitter into two parts. The reflected part was taken as the reference representing the incident light energy and the transmitted beam was focused onto the sample by a focusing mirror (f = 25 cm). Both the incident and transmitted pulse energies were measured simultaneously by two energy detectors (Laser Precision RjP-735). The minimum beam waist of the focused laser beam was 35 or 45 μm, determined by the standard Z-scan method [16]. The optical limiting measurements were conducted with the sample fixed at the focus. To conduct the Z-scans, the sample was moved along the laser light propagation direction while both incident and transmitted laser pulse energies were recorded. In addition, the photoluminescence (PL) measurement for each sample was carried out using a luminescence spectrometer (LS 55, Perkin-Elmer Instrument U.K.) with an excitation wavelength of 442 nm.

## 3. Results and discussion

### 3.1. Optical limiting properties

Figure 2(a) displays the optical limiting results for 1-mm-thick solutions of $C_{60}$, FCA, and FCA/PS composites. Polystyrene in 1,2-dichlorobenzene solution has no optical limiting responses and exhibits 100% transmittance, while $C_{60}$, FCA, and FCA/PS have a linear transmission with input fluences of up to 0.5 J/cm$^2$. When the incident fluence is greater than 0.5 J/cm$^2$, the transmittance begins to decrease and the limiting effect occurs. The optical limiting properties of the FCA itself is poorer than that of its parent $C_{60}$, which may be due to disturbed π-electron system of $C_{60}$ cage upon derivatization. In contrast, the limiting performance of the FCA/PS composites is better than that of $C_{60}$. For $C_{60}$, it is well known that the optical limiting can be described by a five-level model [17,18], in which the limiting behaviour is attributed predominantly to the stronger absorption cross section of the lowest excited triplet state. The better limiting in the FCA/PS composites implies that the triplet-triplet absorption in mono-functional

FCA has been gained by attaching polystyrene. It is obvious that at an input fluence of 1 J/cm$^2$, a strong unstable nonlinear behaviour becomes observable for FCA/PS-B sample. This is probably due to the distortion caused by the interaction between the mono-functional part and polystyrene. However, when the polystyrene concentration is increased from 0.6 to 1.1 g/l in FCA/PS-A, the optical limiting response in the high fluence regime is improved with more reliable behaviour. This indicates that the high polystyrene concentration has marginally effects on the excited triplet state properties of the FCA. Furthermore, the optical limiting of FCA/PS-A has been slightly improved due to the enhancement of delocalised π–electron. Since FCA is an electron-deficient molecule, and PS has a relatively high electron-density, therefore the presence of PS might cause an increase in π–electron density of FCA.

Figure 2(b) more clearly demonstrates the poly (styrene-co-4-vinylpyridiene) concentration dependence of the nonlinear transmission of the FCA/PSVPy32 composites. For the lower PSVPy32 concentration in the composite, the limiting response is nearly the same as that of FCA. However, the higher PSVPy32 concentration added into the composite leads to the better limiting properties. Hence it indicates that the high PSVPy32 concentration gains enhanced excited triplet state properties of FCA, which may be attributed to the enhancement of triplet-triplet state absorption, through increasing triplet-state population, or prolonging triplet-state lifetime of the FCA/PSVPy32.

Table 1. Optical limiting performance measured using 7-ns laser pulses at $\lambda$ = 532 nm

| Sample | Solvent | FCA Concentration (g/l) | Polystyrene Concentration (g/l) | PSVPy32 Concentration (g/l) | Limiting Threshold (J/cm$^2$) | Optical damaged intensity (J/cm$^2$) |
|---|---|---|---|---|---|---|
| $C_{60}$-c | Toluene | - | - | - | 3.1 | 6.9 |
| FCA | 1.2-dichlorobenzene | 1.0 | - | - | 4.1 | 6.4 |
| FCA/PSVPy32-A | 1.2-dichlorobenzene | 1.0 | - | 1.1 | 2.3 | 6.6 |
| FCA/PSVPy32-B | 1.2-dichlorobenzene | 1.0 | - | 0.6 | 5.4 | 6.5 |
| FCA/PS-A | 1.2-dichlorobenzene | 1.0 | 1.1 | - | 2.0 | 6.7 |
| FCA/PS-B | 1.2-dichlorobenzene | 1.0 | 0.7 | - | 2.9 | 6.7 |

If we define the limiting threshold to be the input fluence at which the transmittance falls to 50% of the normalized linear transmittance, Table 1 shows that the limiting threshold of FCA is higher than that of its parent $C_{60}$. However, after FCA have been incorporated with a higher polymer concentration, the limiting threshold of the mono-functional FCA/polymer composites became lower due to a strong interaction between the polymer and FCA. Therefore, the mono-functional FCA/polymer composites are effective in improvement of the optical limiting properties of $C_{60}$. The best limiting performance is found in the FCA/PS-A composite, which offers a limiting threshold 35% lower than that of $C_{60}$.

To check photostability of the samples, we have measured and compared the absorption spectra before and after laser irradiation. The obtained results indicated that there is no difference in the spectra for all the samples. Therefore, all the samples have a good photostability. In addition, the optical damage thresholds of all the samples were in the range of 6.4~6.9 J/cm$^2$, obtained from the optical limiting measurements.

*3.2. Nonlinear optical properties and photoluminescence (PL)*

The nonlinear optical absorption process that contributes to the optical limiting in $C_{60}$ and the FCA/polymers can be explained in terms of the five-level model. For nanosecond laser pulses, transient reverse saturable absorption can be approximately described by a sequential two-photon absorption (STPA), defined as $\alpha = \alpha_0[S_0] + \beta_{eff}I$, where $\beta_{eff} = \alpha_0\alpha_1[S_0]$, $[S_0]$ is the ground state concentration, $\alpha_0$ and $\alpha_1$ are the absorption coefficients describing the ground-state ($S_0$) and triplet-state ($T_1$) absorption, respectively. Such a process can be investigated by the Z-scan method [16].

Figure 3(a) shows the open aperture Z-scans of $C_{60}$, $C_{60}$/PSVPy32, and $C_{60}$/PS in toluene. They have been measured by using 5-ns laser pulses of $\lambda$ = 532 nm at the input irradiance of 160 MW/cm$^2$ (or an input fluence of 0.8 J/cm$^2$). One can conclude that there is no significant difference among these Z-scans, which eliminates the possibility of direct interaction effects between $C_{60}$ and PS or PSVPy32 [19]. However, the open aperture Z-scans are different for $C_{60}$ in toluene, and FCA, FCA/PSVPy32-A, and FCA/PS-A in 1.2-dichlorobenzene, as shown in Figure 3(b). One can see that there is a significant alteration in the nonlinear absorption, in agreement with the limiting measurements in Figure 2. The strongest Z-scan signal is observed in the FCA/PS-A sample. The Z-scan measurements of all the samples were conducted at ~0.8 J/cm$^2$, which is consistent with their optical limiting results. It suggests that the FCA with high concentration of PS/PSVPy32 is a very promising for optical sensor protection. To ensure there is no an existence of small optical nonlinearity contributed by PS or PSVPy32 in 1.2-dichlorobenzene, we have carried out the Z-scan and have observed no nonlinear signal, as shown in Figure 3(b).

To evaluate that there is light emission from the lowest excited singlet state to the ground state, that may affect the population of excited electrons at the excited triplet states, we have performed photoluminescence (PL) measurements at room temperature. Figure 4(a) shows that there is an increase in the PL intensity of mono-functional FCA, in comparison to its parent $C_{60}$. This is due to a significant disturbance of π-electron system of $C_{60}$ cage upon derivatization, consistent with our optical limiting and Z-scan data. Figure 4(b) shows that the presence of polymer, either PSVPy32 or PS, in FCA slightly reduces the PL intensity. Moreover, the PL spectra show that there is polymer concentration dependence: the lower the polymer concentration is, the higher the PL intensity is. This implies an enhancement in the emission from the first excited singlet state to the ground state. Consequently, it decreases the probability of transferring electrons to the lowest excited triplet state by inter-system crossing, leading to poorer optical limiting behaviour. In addition, it should be noted that the PL emission of $C_{60}$ and FCA/polymer composites still exhibits very weak luminescence at room temperature due to high molecular symmetry of $C_{60}$. However, their PL may increase dramatically as they are cooled to low temperature because of reduction of thermal quenching of excited states [20].

A closed relationship between optical limiting and polymer concentration of the mono-functional FCA/polymers indicates that photo-excited electrons and their dynamic process play important role. This is different from the case of polymer/multi-walled carbon nanotube composites, in which polymers have negligible effects because of laser-induced nonlinear scattering - a completely different limiting mechanism [21,22].

**4. Conclusion**

In conclusion, the optical limiting, Z-scan and PL measurements have been carried out to study the nonlinear optical properties of mono-functional FCA/polymer composites. The FCA/polymer composites possess better optical limiting properties than that of FCA, which are mainly contributed by stronger absorption of excited triplet state due to the presence of high concentration polystyrene or PSVPy32.

**Acknowledgements**

We thank the National University of Singapore for the financial support of this work.

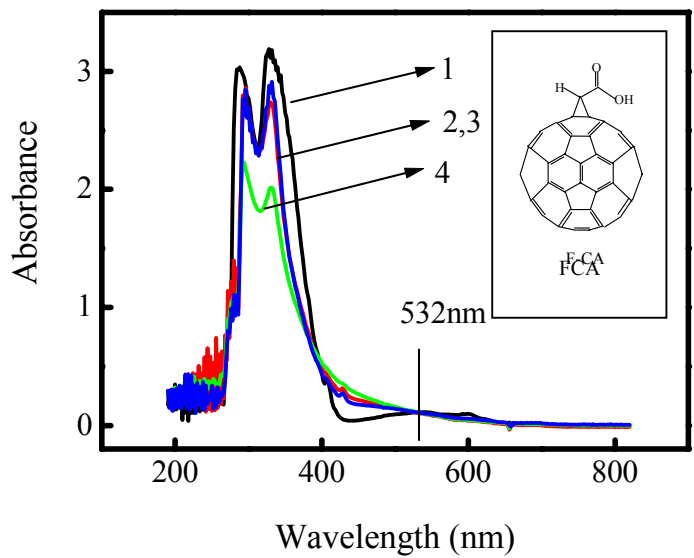

**Fig. 1.** UV-Vis absorption spectra of 1: $C_{60}$ in toluene, 2: FCA, 3: FCA/PS-A, and 4: FCA/PSVPy-A, wherein, 2, 3 and 4 are in 1,2-dichlorobenzene solutions. The inset is the mono-functional FCA structure. All the solutions are directly used in the optical limiting and Z-scan measurements.

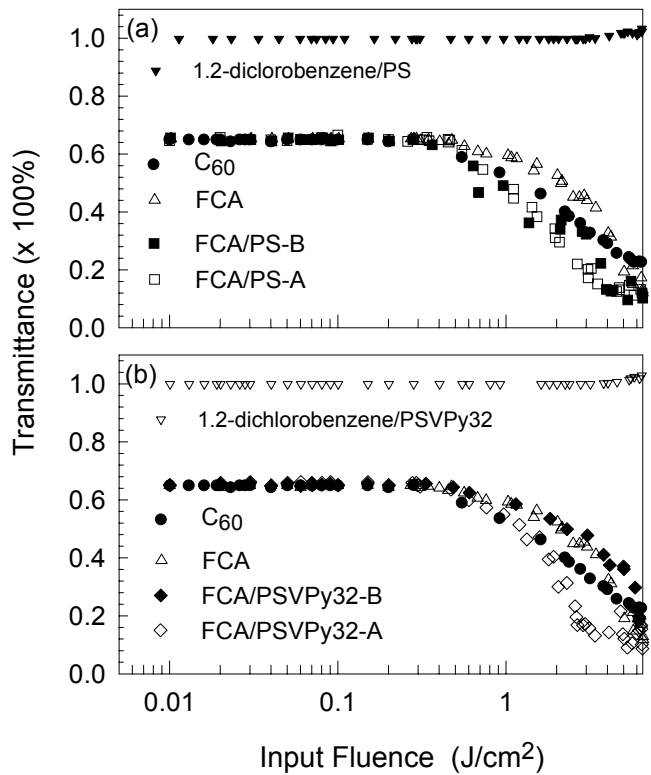

**Fig. 2. (a)** Nonlinear transmission of $C_{60}$-toluene (filled circles), FCA in 1.2-dichlorobenzene (open triangles), FCA/PS-B in 1.2-dichlorobenzene (filled squares), FCA/PS-A in 1.2-dichlorobenzene (open squares), and PS in 1.2-dichlorobenzene (filled inverted triangles). **(b)** Nonlinear transmission of $C_{60}$-toluene (filled circles), FCA in 1.2-dichlorobenzene (open triangles), FCA/PSVPy32-B in 1.2-dichlorobenzene (filled diamonds), FCA/PSVPy32-A in 1.2-dichlorobenzene (open diamonds), and PSVPy32 in 1.2-dichlorobenzene (open inverted triangles). The measurements are conducted with 7-ns laser pulses at $\lambda = 532$ nm. All the samples have the same linear transmission of $T = 65\%$ and the optical path of 1 mm.

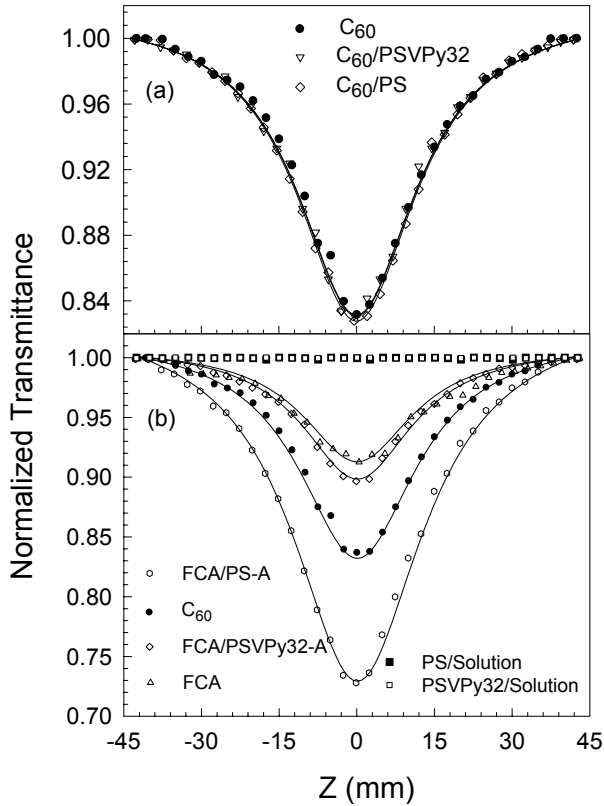

**Fig. 3.** **(a)** Open aperture Z-scans conducted at the input irradiance $I$ = 160 MW/cm$^2$ and beam waist of 45 μm on 1-mm-thick solutions of $C_{60}$ (filled circles), $C_{60}$/PS (open inverted triangles) and $C_{60}$/PSVPy32 (open diamonds) in toluene. **(b)** Open aperture Z-scans performed with $I$ = 160 MW/cm$^2$ on 1-mm-thick solutions of $C_{60}$ in toluene (filled circles), FCA in 1.2-dichlorobenzene (open triangles), FCA/PSVPy32-A in 1.2-dichlorobenzene (open diamonds), FCA/PS-A in 1.2-dichlorobenzene (open hexagons), PS in 1.2-dichlorobenzene (filled squares), and PSVPy32 in 1.2-dichlorobenzene (open squares). The measurements are conducted with 5-ns laser pulses at λ = 532 nm. The solid curves are the best fits by the Z-scan theory [16].

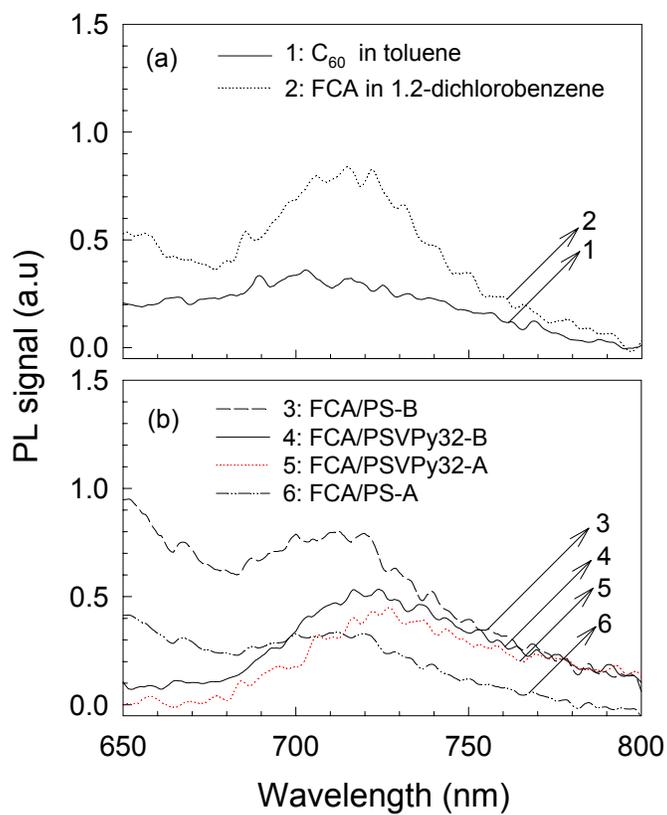

**Fig. 4.** **(a)** Photoluminescence (PL) spectra of 10-mm-thick 1: $C_{60}$ in toluene, and 2: FCA in 1,2-dichlorobenzene solution. **(b)** PL spectra of 10-mm-thick 3: FCA/PS-B, 4: FCA/PSVPy32-B, 5: FCA/PSVPy32-A and 6: FCA/PS-A in 1,2-dichlorobenzene, respectively.